\documentclass[preprint]{elsarticle}
\usepackage{graphicx}          % recommended for figures
\usepackage{amsmath, amsbsy, amssymb}
\usepackage{hyperref}

\begin{document}

\begin{frontmatter}
 \title{Simulation of Multivariate Non-Gaussian Autoregressive Time Series with Given Autocovariance and Marginals}

 \author[DK]{Dimitris~Kugiumtzis\corref{cor1}}
 \ead{dkugiu@auth.gr}
 \ead[url]{http://users.auth.gr/dkugiu}
 
 \author[EB]{Efthimia~Bora-Senta}
 \ead{bora@math.auth.gr}

 \cortext[cor1]{Corresponding author}
 \address[DK]{Department of Electrical and Computer Engineering, Faculty of Engineering, Aristotle University of Thessaloniki, Thessaloniki 54124, Greece}
 \address[EB]{Department of Mathematics, Aristotle University of Thessaloniki, Thessaloniki 54124, Greece}

\begin{abstract}
A semi-analytic method is proposed for the generation of realizations of a multivariate process of a given linear correlation structure and marginal distribution. This is an extension of a similar method for univariate processes, transforming the autocorrelation of the non-Gaussian process to that of a Gaussian process based on a piece-wise linear marginal transform from non-Gaussian to Gaussian marginal. The extension to multivariate processes involves the derivation of the autocorrelation matrix from the marginal transforms, which determines the generating vector autoregressive process. The effectiveness of the approach is demonstrated on systems designed under different scenarios of autocovariance and marginals.

\end{abstract}


\end{frontmatter}

% {\em Running title:} Simulation of Multivariate Non-Gaussian }

% \date{}

% \maketitle

% \bibliographystyle{authordate1}
\bibliographystyle{elsarticle-num}

%%%%%%%%%%%%%%%%%%%%%%%%%%%%%%
\section{Introduction}
\label{sec:Intro}
%%%%%%%%%%%%%%%%%%%%%%%%%%%%%%

Many real time series cannot be considered to be Gaussian and do not fit in the framework of standard linear analysis. A related problem is the generation of non-Gaussian time series with given linear correlation structure and marginal distribution. This problem arises mainly in stochastic simulation and randomization testing. 

In stochastic simulation and particularly in the framework of input modelling, the problem occurs when there are dependencies among random variables that constitute the inputs to the simulation model. For time series data, this leads to the simulation of univariate or multivariate stochastic linear processes of given marginal distributions and lagged correlation structure \citep{Cario96,Biller05,Kuhl10,Biller11}. This setting is met in many applications, ranging from manufacturing systems \citep{Ip99}, to medical treatment \citep{He12}, internet traffic \citep{Kriege11}, bird flocking \citep{Schruben10}, floods \citep{Cai11}, and ocean temperature \citep{Das09}. 

Randomization testing has been used to investigate nonlinear dependencies in the time series, where the null hypothesis is that the underlying process is linear stochastic and the test statistic is nonlinear. For many nonlinear statistics the null distribution is not known and a common approach, known as surrogate data test, is to form the empirical null distribution from the values of the test statistic computed on randomized time series consistent to the null hypothesis. The latter requires that the surrogate time series preserve the original marginal distribution and linear correlation structure \citep{Theiler92,Schreiber96,Schreiber99b,Kugiumtzis02}. The surrogate data test for nonlinearity has been mainly developed for univariate time series and has been applied in many fields for the investigation of nonlinear dynamics and chaos, such as finance \citep{Kugiumtzis01a,Small03,Das07}, geophysics \citep{Pavlos99,Pavlos99b,Keylock10} and physiology \citep{Pradhan96,Mormann05}. For multivariate time series, there are few approaches approximating the linear correlation structure in the frequency domain using the cross-power spectrum  \citep{Dolan02,Andrzejak03b,Faes10}.  

Though the problem is the same, solutions were proposed independently in the two areas of stochastic stimulation and surrogate data testing. In stochastic simulation, the problem was postulated as generating time series of arbitrary length with a given autocorrelation and marginal distribution. In a series of works, the algorithm for univariate time series called autoregressive-to-anything (ARTA) was developed, modified and tested \citep{Li75,Cario96,Cario98,Biller08}, and was further extended for multivariate time series, known as Vector ARTA (VARTA) \citep{Biller03}. The method relies on solving numerically a double integral expression for the transform of the product moment of two Gaussian variables to that of variables of arbitrary marginals, the latter being the lagged variables of the linear stochastic process of arbitrary marginal distribution. The computations may be simplified approximating the marginal distribution with the Johnson translation system of distributions \citep{Biller03}, and the generalized Pareto distribution \citep{Cai11}. 

For the surrogate data test for nonlinearity, independently of the ARTA approaches, different algorithms were developed to generate time series that match the marginal distribution and correlation structure of the examined time series. All these methods use data randomization, matching exactly the sample marginal distribution, and approximating the linear correlation either in the frequency domain using the Fourier transform \citep{Theiler92,Schreiber96}, refined further using wavelet transform \citep{Keylock07}, or in the time domain, correcting the autocorrelation function \citep{Kugiumtzis00d}, or finding an appropriate Gaussian autoregressive process, called statically transformed autoregressive process (STAP) \citep{Kugiumtzis02}. 
STAP and ARTA are similar in that both methods attempt to form the transform from Gaussian autocorrelation to the given autocorrelation, but ARTA uses numerical solution of the double integral form, while STAP uses parametric approximation, originally polynomial \citep{Kugiumtzis02}, and then piece-wise linear \citep{Kugiumtzis10}.  

Here, we extend the piece-wise approximation in STAP to estimate the auto- and cross-correlation of the multivariate time series, and we term the method as vector STAP (VSTAP). We demonstrate the performance of VSTAP on different simulated multivariate stochastic processes.

The structure of the paper is as follows. In Section~\ref{sec:STAP}, we give the background and briefly discuss STAP using piece-wise approximation for univariate time series, and in Section~\ref{sec:VSTAP}, we present VSTAP for multivariate time series. In Section~\ref{sec:Simulations}, we show the results of simulations on different multivariate stochastic processes, and we conclude in Section~\ref{sec:Conclusion}. 

%%%%%%%%%%%%%%%%%%%%%%%%%%%%%%
% \section{The Method of Statically Transformed Autoregressive Process}
\section{Univariate Time Series with given Marginal Distribution and Autocorrelation}
\label{sec:STAP}
%%%%%%%%%%%%%%%%%%%%%%%%%%%%%%

We start with the univariate case, and suppose it is given a univariate time series $\{x_t\}_{t=1}^{n}$ with marginal distribution $F_{X}(x)$ and sample autocorrelation function $r_X(\tau)$, where $\tau$ is the time lag. Equivalently, instead of $r_X(\tau)$ the power spectrum $S_X(f)$ may be considered, where $f$ is the frequency. The problem is to generate a time series $\{x^*_t\}_{t=1}^{N}$, where $N$ may be different from $n$, fulfilling the following two conditions:
\begin{align}
F_{X^*}(x) & = F_{X}(x) 
\label{eq:uniCon1} 
\\
r_{X^*}(\tau) & = r_X(\tau), \quad \tau=1,\ldots,P
\label{eq:uniCon2} 
\end{align}
for a sufficiently large $P$. The second condition for the preservation of the linear correlation structure can be equivalently given in terms of power spectrum, $S_{X^*}(f) = S_X(f)$ for all frequencies $f$. In stochastic simulation, the problem may be postulated without reference to a specific time series $\{x_t\}_{t=1}^{n}$ but only to the given $F_{X}(x)$ and $r_X(\tau)$. 

% --------------------------
\subsection{Proposed solutions}
\label{subsec:STAPsolutions}
% --------------------------

The solutions we consider here match exactly the first condition in eq.(\ref{eq:uniCon1}) and approximate the second condition in eq.(\ref{eq:uniCon2}). All solutions make use of the marginal transform from Gaussian to the given distribution 
\begin{equation}
   x = F_X^{-1}(\Phi(z))
   \label{eq:ztox}
\end{equation}
and the inverse transform 
\begin{equation}
z = \Phi^{-1}(F_X(x)), 
\label{eq:xtoz}
\end{equation}
assuming a variable $Z$ following the standard Gaussian distribution with cumulative density function (cdf) $\Phi$. 

There are two main approaches for the solution: the constrained realization approach, where the objective is to transform a random time series in order to match the given two conditions, and the typical realization approach, attempting to find a generating process that fulfills the two conditions. Though there has been some evidence in favor of constrained realization for hypothesis testing, as is the case with the surrogate data test for nonlinearity \citep{Theiler96b}, it requires a time series $\{x_t\}_{t=1}^{n}$ is given, from which the marginals and correlation structure are derived, and it gives another time series $\{x^*_t\}_{t=1}^{n}$ of the same length. On the other hand, the typical realization approach gives more insight onto the underlying process and has also the advantage that it can take as input only the marginals and autocorrelations, and can generate time series of any length, e.g. see \cite{Halley11}. The amplitude adjusted Fourier transform (AAFT) \citep{Theiler92} and the iterated AAFT (IAAFT) \citep{Schreiber96} transform a time series to approximate the power spectrum and are constrained realization approaches. On the other hand, ARTA and STAP attempt to identify the process that generates realizations possessing the given autocorrelation and are therefore typical realization approaches. The two latter approaches are decomposed in the same four steps:
\begin{enumerate}
    \item Starting with the marginal transform in eq.(\ref{eq:ztox}), the transform $\psi_{\tau}$ from the Gaussian autocorrelation $r_Z(\tau)$ to the given autocorrelation $r_X(\tau)$ is determined for each lag $\tau$, $r_X(\tau) = \psi_{\tau} (r_Z(\tau))$, and the solution for each $r_Z(\tau)$ is obtained. 
    
    \item For a given order $P$, the coefficients of an AR($P$) process are computed from the autocorrelations $r_Z(\tau)$, $\tau=1,\ldots,P$, using the Yule-Walker equations \citep[Sec. 7.1]{Wei06}.

    \item A Gaussian time series $\{z^*_t\}_{t=1}^N$ of a given length $N$ is generated by the AR($P$) process.

    \item The Gaussian time series is transformed to obtain the given marginal distribution, $x_t^* = F^{-1}(\Phi(z^*_t))$, resulting in the desired time series $\{x^*_t\}_{t=1}^{N}$. 
\end{enumerate}

% --------------------------
\subsection{The method of statically transformed autoregressive process}
\label{subsec:STAPmethod}
% --------------------------

ARTA and STAP differ only in the first step. In ARTA, the double integral form for $\psi_{\tau}$ is derived based on the marginal transform \citep{Cario96} 
\begin{align}
    r_X(\tau) & = \mbox{Corr}(F_X^{-1}(\Phi(z_t)),F_X^{-1}(\Phi(z_{t-\tau}))) \nonumber \\ 
               & = \frac{1}{s^2}\left(\int_{-\infty}^{\infty}\int_{-\infty}^{\infty} F_X^{-1}(\Phi(z_t)) F_X^{-1}(\Phi(z_{t-\tau}))\phi(z_t,z_{t-\tau},r_{Z}(\tau)) \mbox{d}z_t \mbox{d}z_{t-\tau} - \bar{x}^2\right),
\label{eq:ARTA}
\end{align}
where $\phi(z_t,z_{t-\tau},r_{Z}(\tau))$ is the bivariate standard Gaussian probability density function (pdf), and 
$\bar{x}$ and $s^2$ are the sample mean and variance of $X$, respectively. To solve eq.(\ref{eq:ARTA}) with respect to $r_{Z}(\tau)$, the double integral form is solved numerically (for an enhanced numerical solution, see \citep{Chen01}). 

On the other hand, STAP uses a parametric approximation of $\psi_{\tau}$. The original STAP in \citep{Kugiumtzis02} uses polynomial approximation of the marginal transform in eq.(\ref{eq:ztox}), resulting in a polynomial form for $\psi_{\tau}$. We found that linear piece-wise approximation gives a better solution of $r_X(\tau) = \psi_{\tau} (r_Z(\tau))$ with respect to $r_Z(\tau)$ \citep{Kugiumtzis10}. The linear piece-wise approximation of the marginal transform is comprised of first degree polynomials at each of $m$ segments defined by $m-1$ breakpoints $a_k$, $k=1,\ldots,m-1$ 
\begin{equation}
X_t = \left\{\begin{array}{c@{\;\;\;\mbox{if}\;\;\;}c}
 c_{10}+c_{11}Z_t & -\infty<Z_t\le a_1 \\
 c_{20}+c_{21}Z_t & a_1<Z_t\le a_2 \\
 \multicolumn{2}{c}{\dotfill} \\
 c_{m0}+c_{m1}Z_t & a_{m-1}<Z_t<\infty 
 \end{array} \right.
\label{eq:pieceline}
\end{equation}
The partition of $Z$ and the linear piece-wise function in
(\ref{eq:pieceline}) determine a partition $\{A_k \, | \, k=1,\ldots,m\}$ of
the domain of $X$, i.e. $X \in A_k$ when $Z \in [a_{k-1},a_k]$. Specifically, we have $A_k= [\alpha_{k-1},\alpha_k] = [F_X^{-1}(\Phi(a_{k-1})),F_X^{-1}(\Phi(a_{k}))]$. 
Then the product moment for a lag $\tau$ is 
\begin{align}
\mbox{E}(X_t X_{t-\tau}) & = \sum_{k=1}^m \sum_{l=1}^m \mbox{E}(X_t X_{t-\tau} | X_t \!\in\! A_k \wedge X_{t-\tau} \!\in\! A_l) \mbox{Pr}(X_t \!\in\! A_k \wedge X_{t-\tau} \!\in\! A_l) \nonumber \\
& = \sum_{k=1}^m \sum_{l=1}^m
(c_{k0}c_{l0}+c_{k1}c_{l0}\mu_{1,0}+c_{k0}c_{l1}\mu_{0,1}+c_{k1}c_{l1}\mu_{1,1})P,
\label{eq:piecemoment}
\end{align}
where $P=P(a_{k-1},a_k,a_{l-1},a_l;\rho_Z(\tau))$ is the probability of $(Z_t,Z_{t-\tau})$ being in the region $[a_{k-1},a_k]\times [a_{l-1},a_l]$, $\mu_{1,0}$ and $\mu_{0,1}$ are the first order marginal moments and $\mu_{1,1}$ the product moment of the doubly truncated Gaussian variables $(Z_t,Z_{t-\tau})$, where $Z_t \in [a_{k-1},a_k]$ and $Z_{t-\tau} \in [a_{l-1},a_l]$. Substituting the expressions for the moments of the joint doubly truncated Gaussian distribution, we get an analytic form for $\psi_{\tau}$ (for details, see \citep{Kugiumtzis10}). We found that best results are obtained when the breakpoints divide the standard Gaussian domain into equiprobable intervals. We also found that the constraint of continuity on the linear piece-wise function does not affect substantially the approximation of the correlation transform. Therefore we independently estimate the coefficients of the linear function at each interval rather than using linear splines. It is noted that the possible violation of continuity does not affect the monotonicity of the piece-wise linear function, which is always  maintained.

In a comparative study in \citep{Kugiumtzis08a}, it was shown that $r_{X^*}(\tau)$ from STAP with the polynomial approximation estimates $r_{X}(\tau)$ without bias, as opposed to AAFT, IAAFT and a model bootstrap approach, but with a larger variance than IAAFT, which decreases with the increase of $n$. The linear piecewise approximation decreases further the variance of $r_{X^*}(\tau)$ from STAP and simulations in \citep{Kugiumtzis10} showed that it gives more accurate Gaussian correlation estimation, and therefore we adopt it in the extension of STAP for multivariate time series presented below.

%%%%%%%%%%%%%%%%%%%%%%%%%%%%%%
\section{The Method of Vector Statically Transformed Autoregressive Process}
\label{sec:VSTAP}
%%%%%%%%%%%%%%%%%%%%%%%%%%%%%%

For $K$ multivariate time series, the problem involves the marginal distributions of all variables $X_1,\ldots,X_K$, and the lagged cross-correlation for all pairs $(X_i,X_j)$ in addition to the autocorrelations for each $X_i$.
Given the marginal distributions $F_{X_i}(x)$, $i=1,\ldots,K$, and lagged correlations $r_{X_i,X_j}(\tau)$, the problem is to find a multivariate time series $\{x^*_{i,t}\}_{t=1}^{N}$, $i=1,\ldots,K$, fulfilling the two following conditions:
\begin{align}
F_{X_i^*}(x) & = F_{X_i}(x), \quad i=1,\ldots,K
\label{eq:multiCon1} 
\\
r_{X_i^*,X_j^*}(\tau) & = r_{X_i,X_j}(\tau), \quad \tau=0,\ldots,P, \quad i,j=1,\ldots,K
\label{eq:multiCon2} 
\end{align}
for a sufficiently large $P$. Alternatively, considering the linear structure in the frequency domain the second condition can be postulated in terms of cross-power spectrum, $S_{X_i^*,X_j^*}(f) = S_{X_i,X_j}(f)$ for all frequencies $f$. 

The first condition on the marginal distributions does not really complicate the solution for a proper $\{x^*_{i,t}\}_{t=1}^{N}$ as the marginal transform can be applied separately for each of the $K$ variables. On the other hand, the preservation of the linear correlation structure is far more difficult to achieve than for the univariate case because in addition to the autocorrelation $r_{X_i,X_i}(\tau)$ (denoted $r_X(\tau)$ in the univariate case) the lagged cross-correlations $r_{X_i,X_j}(\tau)$, $i\neq j$, have to be matched as well (the same holds for the cross-power spectrum). 

% --------------------------
\subsection{The typical realization approach}
\label{subsec:VSTAPsolution}
% --------------------------

The typical realization approach is similar to that for univariate time series and is decomposed in the following four steps:
\begin{enumerate}
    \item The correlation transform for each pair of variables $(X_i,X_j)$ and lag $\tau$ is formed as $r_{X_i,X_j}(\tau) = \psi_{i,j,\tau}(r_{Z_i,Z_j}(\tau))$ from the marginal transforms, and the solution for each $r_{Z_i,Z_j}(\tau)$ is obtained.
\label{step:VSTAP1}
    
    \item Given the lagged cross- and auto-correlations $r_{Z_i,Z_j}(\tau)$, $i,j=1,\ldots,K$, $\tau=0,\ldots,P$ (for a given order $P$), the coefficient matrices of a vector autoregressive process of order $P$ on $K$ variables, VAR$_K(P)$, are computed  using the multivariate generalization of the Yule-Walker equations \citep[Sec. 16.5]{Wei06}.
\label{step:VSTAP2}

    \item The Gaussian multivariate time series $\{z^*_{i,t}\}_{t=1}^{N}$, $i=1,\ldots,K$ is generated by the VAR$_K(P)$ process.
\label{step:VSTAP3}

    \item Each of the Gaussian time series $\{z^*_{i,t}\}_{t=1}^{N}$ is transformed to obtain the given marginal distribution, $x_{i,t}^* = F_{X_i}^{-1}(\Phi(z^*_{i,t}))$, resulting in the desired time series $\{x^*_{i,t}\}_{t=1}^{N}$.
\label{step:VSTAP4}
\end{enumerate}

Albeit the similarity of the four steps above to these for the univariate case in Sec.~\ref{sec:STAP}, there are additional problems in their implementation, which will be discussed later in Sec.~\ref{subsec:VSTAPproblems}. Both the vector ARTA (VARTA)
\citep{Biller03}, and vector STAP (VSTAP), presented below, implement the four steps for any marginal distributions and cross-correlation structure. Further, both methods can be implemented to multivariate time series assuming only that they are continuous valued and stationary. The main difference in VSTAP and VARTA is that in step~\ref{step:VSTAP1}, VARTA derives $r_{Z_i,Z_j}(\tau)$ solving numerically the double integral form as in eq.(\ref{eq:ARTA}), while 
VSTAP approximates the marginal transform from Gaussian to the given sample distribution with a linear piece-wise function, given in eq.(\ref{eq:pieceline}).

\subsection{Implementation of VSTAP}

VSTAP first fits the linear piece-wise function to the sample marginal transform, as for the univariate case (see eq.(\ref{eq:pieceline})). The breakpoints of the piece-wise function divide the standard Gaussian domain to equiprobable intervals and are thus the same for all variables, and only the coefficients of the linear piece-wise function are different for each variable $Z_i$, denoted $c_{i,k0}$ and $c_{i,k1}$, $k=1,\ldots,m$, for the constant term and the slope coefficient, respectively. The product moment for the pair of variables $(X_i,X_j)$ and lag $\tau$ is similar to that of eq.(\ref{eq:piecemoment}) for the univariate case
\begin{align}
\mbox{E}(X_{i,t} X_{j,t-\tau}) & = \sum_{k=1}^m \sum_{l=1}^m \mbox{E}(X_{i,t} X_{j,t-\tau} | X_{i,t} \!\in\! A_{i,k} \wedge X_{j,t-\tau} \!\in\! A_{j,l}) \mbox{Pr}(X_{i,t} \!\in\! A_{i,k} \wedge X_{j,t-\tau} \!\in\! A_{j,l}) \nonumber \\
& = \sum_{k=1}^m \sum_{l=1}^m
(c_{i,k0}c_{j,l0}+c_{i,k1}c_{j,l0}\mu_{1,0}+c_{i,k0}c_{j,l1}\mu_{0,1}+c_{i,k1}c_{j,l1}\mu_{1,1})P,
\label{eq:vpiecemoment}
\end{align}
where $P=P(a_{k-1},a_{k},a_{l-1},a_{l};\rho_{Z_i,Z_j}(\tau))$ is the probability of $(Z_{i,t},Z_{j,t-\tau})$ being in the region $[a_{k-1},a_{k}]\times [a_{l-1},a_{l}]$, $\mu_{1,0}$ and $\mu_{0,1}$ are the first order marginal moments and $\mu_{1,1}$ the product moment of the doubly truncated bivariate standard Gaussian distribution on $[a_{k-1},a_{k}]\times [a_{l-1},a_{l}]$. 
The first order marginal moments in eq.(\ref{eq:vpiecemoment}) have the same expression for all $i=1,\ldots,m$, e.g. $\mu_{1,0}$ is
\[
\mu_{1,0} = \mbox{E}(Z_{i,t}) = \frac{1}{P}\sum_{u,v=0}^1
(-1)^{u+v}\left(\phi(a_{k-u})Q(a_{k-u},a_{l-v})+\rho \phi(a_{l-v})Q(a_{l-v},a_{k-u})\right),
\] 
where $Q(a,b) = \int_{\frac{b-\rho a}{\sqrt{1-\rho^2}}}^{\infty}\varphi(u)\,\mathrm{d}u$ and $\rho=\rho_{Z_i,Z_j}(\tau)$. The product moment in eq.(\ref{eq:vpiecemoment}) is  
\begin{align*}
\mu_{1,1} = \mbox{E}(Z_{i,t},Z_{j,t-\tau}) = & \frac{1}{P}\sum_{u,v=0}^1
(-1)^{u+v}\left(\rho P+(1-\rho^2)\phi(a_{k-u},a_{l-v},\rho) \right.\nonumber \\
& \left.+\rho a_{k-u}\phi(a_{k-u})Q(a_{k-u},a_{l-v})+\rho a_{l-v}\phi(a_{l-v})Q(a_{l-v},a_{k-u})\right).
\end{align*}
The probability $P$ and the moments are functions of $\rho_{Z_i,Z_j}(\tau)$, so after substitution we get an analytic form for $\hat{\psi}_{i,j,\tau}$ that approximates the true $\psi_{i,j,\tau}$ based on linear piece-wise marginal transforms \citep{Kugiumtzis10}. The function $\hat{\psi}_{i,j,\tau}$ is invertible, given that the piece-wise approximation in eq.(\ref{eq:pieceline}) is also invertible \citep[Theorem 3.4]{Biller03} (the theorem is based on the monotonicity of the cdf $F_{X_i}$, and thus it applies to the piece-wise function fitted to $F_{X_i}$ being monotonic). However, the expression of $\hat{\psi}_{i,j,\tau}$ is very complicated and the closed form solution for its inverse cannot be obtained. We therefore use an iterative process to obtain the solution $r_{Z_i,Z_j}(\tau)$, approximating $\rho_{Z_i,Z_j}(\tau)$, so that $\hat{\psi}_{i,j,\tau}(r_{Z_i,Z_j}(\tau))$ matches $r_{X_i,X_j}(\tau)$ at an arbitrary accuracy $\epsilon$.  

The iterative process needs a starting value for $r_{Z_i,Z_j}(\tau)$, and an appropriate value is given by the so-called naive correlation coefficient of $X_{i,t}$ and $X_{j,t-\tau}$.  This is the Pearson correlation coefficient of the marginal transforms to Gaussian of $X_{i,t}$ and $X_{j,t-\tau}$ according to eq.(\ref{eq:xtoz}), $r_{Z_i,Z_j}(\tau)^{(0)} = r\left(\Phi^{-1}(F_{X_i}(X_{i,t})),\Phi^{-1}(F_{X_j}(X_{j,t-\tau})) \right)$ \citep{Zou02}. 
The main steps of the iterative algorithm are the following:
\begin{enumerate}
    \item Begin with the naive correlation coefficient of $X_{i,t}$ and $X_{j,t-\tau}$ as the starting value $r_{Z_i,Z_j}(\tau)^{(0)}$.
    \label{step:cortrans1}
    \item At each iteration $h$, compute $r_{X_i,X_j}(\tau)^{(h)} = \hat{\psi}_{i,j,\tau}\left(r_{Z_i,Z_j}(\tau)^{(h)}\right)$.
    \label{step:cortrans2}
    \item Compute the difference $\delta r_{X_i,X_j}(\tau)=r_{X_i,X_j}(\tau)-r_{X_i,X_j}(\tau)^{(h)}$.
    \label{step:cortrans3}
    \item If $\delta r_{X_i,X_j}(\tau)<\epsilon$ the solution is found and $r_{Z_i,Z_j}(\tau)=r_{Z_i,Z_j}(\tau)^{(h)}$. Otherwise the input for the next step is updated as 
$r_{Z_i,Z_j}(\tau)^{(h+1)} = r_{Z_i,Z_j}(\tau)^{(h)} + \delta r_{X_i,X_j}(\tau)$ and the computations are repeated from step~\ref{step:cortrans2}.
    \label{step:cortrans4}
\end{enumerate}
The iterative algorithm is of the type of simple fixed point iteration, and thus has a linear rate of convergence. It requires only one starting value as opposed to bracketing methods (e.g. false position, secant, and bisection) requiring two starting values close and at each side of the solution \citep[Chp.2]{Burden05}. The starting value of the naive correlation coefficient is close to the desired solution \citep{Zou02}, and actually coincides with it in the case ($X_{i,t}, X_{j,t-\tau})$ is obtained from monotonic marginal transforms of a bivariate Gaussian variable pair \citep{Kugiumtzis10}. 
At each iteration, we set the increment in the Gaussian correlation equal to the deviation in the target correlation $\delta r_{X_i,X_j}(\tau)$ because the Gaussian correlation and the target correlation are at the same amplitude level, and moreover it holds $r_{X_i,X_j}(\tau) \le |r_{Z_i,Z_j}(\tau)|$ \citep[p.600]{Kendall79}. The linear convergence is guaranteed by the monotonicity of $\psi_{i,j,\tau}$ and that $r_{X_i,X_j}(\tau) \le |r_{Z_i,Z_j}(\tau)|$. In practice, we have found that the monotonicity of $\hat{\psi}_{i,j,\tau}$ may not hold at the edges of the interval $[-1,1]$. For this, when $|r_{Z_i,Z_j}(\tau)^{(h)}|$ is larger than a threshold close to one (we set the threshold to 0.9 to be on the safe side for all practical purposes) the condition of monotonicity is checked and if it is not satisfied, binary search is applied in the interval formed by the threshold and the current value. 
The number of iterations (including the binary search) depends on the given accuracy $\epsilon$, but in any case the closed form expression of $\hat{\psi}_{i,j,\tau}(r_{Z_i,Z_j}(\tau))$ makes the algorithm very time effective.  

% ----------------------------------------
\subsection{Implementation complications}
\label{subsec:VSTAPproblems}
% ----------------------------------------

A known problem with any correlation transform is that there may not be a feasible solution for a particular given correlation and marginals \citep{Li75,Ghosh02}. The domain of $r_{X_i,X_j}(\tau)$ for which a solution $r_{Z_i,Z_j}(\tau) \in [-1,1]$ can be obtained is a subset of $[-1,1]$. For a bivariate sample $\{x_t,y_t\}_{t=1}^n$ of $(X,Y)$ (in our case $X=X_{i,t}$ and $Y=X_{j,t-\tau}$), the subset is formed by the minimum feasible correlation $\underline{r}$ and the maximum feasible correlation $\overline{r}$
\[
\underline{r} = \frac{\left(\sum_{t=1}^n x_{(t)}y_{(n-t+1)}-\bar{x}\bar{y}\right)/(n-1)}{s_X s_Y} \quad\quad
\overline{r} = \frac{\left(\sum_{t=1}^n x_{(t)}y_{(t)}-\bar{x}\bar{y}\right)/(n-1)}{s_X s_Y},
\]
where $\{x_{(t)}\}_{t=1}^n$ is the ordered sample of $X$, $\bar{x}$ and $s_X$ are the sample mean and standard deviation \citep{Whitt76}. For correlation matrices it is more difficult to determine the condition for feasibility, but there are computational procedures to check whether a given correlation matrix is feasible \citep{Ghosh02}. In the case of multivariate time series, the check for feasibility extends to the correlation matrix of any vector variable with components from the set $\{X_{i,t-\tau} \,\, | \,\, i=1,\ldots,K, \,\, \tau=0,\ldots,P\}$. Even if all the correlation matrices are feasible, the solution for the corresponding Gaussian correlation matrices may not be valid, i.e. the matrices may not all be positive semi-definite. The problem arises because the components of each Gaussian correlation matrix are computed independently. 

Instead of checking positive semidefiniteness for each possible Gaussian correlation matrix, the validity of the derived Gaussian lagged correlations can be tested collectively by checking for positive semidefiniteness of the full Gaussian correlation matrix
\begin{equation}
R_Z=\left[ \begin{array}{cccc}
R_Z(0) & R_Z(1) & \cdots & R_Z(P) \\
R_Z(-1) & R_Z(0) & \cdots & R_Z(P-1) \\
\vdots & \vdots & \vdots & \vdots \\
R_Z(-P) & R_Z(-P+1) & \cdots & R_Z(0) 
\end{array} \right]
\label{eq:AllCrossCorrM}
\end{equation}
where $R_Z(\tau)$ is the lagged correlation matrix for lag $\tau$, $\tau=0,\ldots,P$ ($P$ is the maximum lag)
\begin{equation}
R_Z(\tau)=\left[ \begin{array}{cccc}
1 & r_{Z_1,Z_2}(\tau) & \cdots & r_{Z_1,Z_K}(\tau) \\
r_{Z_2,Z_1}(\tau) & 1 & \cdots & r_{Z_2,Z_K}(\tau) \\
\vdots & \vdots & \vdots & \vdots \\
r_{Z_K,Z_1}(\tau) & r_{Z_K,Z_2}(\tau) & \cdots & 1 
\end{array} \right] 
\label{eq:CrossCorrM}
\end{equation}

For correlation matrices, there are techniques to modify the matrix so as to be positive semidefinite, such as replacing negative eigenvalues with zero, or better with a slightly positive value in order to make it positive definite \citep{Higham02,Ghosh02}. However, these techniques cannot be applied directly to the correlation matrix $R_Z$ in eq.(\ref{eq:AllCrossCorrM}) because it is comprised of repeated blocks being the lagged correlation matrices in eq.(\ref{eq:CrossCorrM}). Thus a change in an eigenvalue of $R_Z$ will alter its structure and make previously identical blocks differ. Our solution to this problem is to introduce a two-stage iterative procedure, where in the first stage we render positive definiteness of $R_Z$ and in the second stage we regain the structure in eq.(\ref{eq:AllCrossCorrM}). At each iteration, in the first stage we set the negative or zero eigenvalues of $R_Z$ to a slightly positive value and obtain a positive definite matrix but with altered components. In the second stage, for each repeated component in the form in eq.(\ref{eq:AllCrossCorrM}), we take the average of the values at the corresponding entries. For example, $r_{Z_1,Z_2}(1)$ occurs in the entry $(1,2)$ of $R_Z(1)$ and entry $(2,1)$ of $R_Z(-1)$, so it occurs in all blocks $R_Z(1)$ and $R_Z(-1)$ of $R_Z$, i.e. $2(P-1)$ entries in total. Replacing all the repeated entries with the same average value gains back the correct structure of $R_Z$ but may cause $R_Z$ not to be positive definite, and the same two-stage procedure is then repeated. We have not worked out a proof for the convergence of this iterative procedure, but we found that a positive definite matrix $R_Z$ of the form in eq.(\ref{eq:AllCrossCorrM}) could be obtained after few steps.

According to step~\ref{step:VSTAP2} of the typical realization approach in Sec.~\ref{subsec:VSTAPsolution}, given $R_Z$ the coefficient matrices $A_1,\ldots,A_P$ of the VAR$_K(P)$ Gaussian process $\{Z_{1,t},\ldots,Z_{K,t}\}$ are computed from the multivariate generalization of the Yule-Walker equations \citep[Sec. 16.5]{Wei06}, where VAR$_K(P)$ is expressed as 
\begin{equation}
\mathbf{Z}_t = A_1 \mathbf{Z}_{t-1} + \cdots A_P \mathbf{Z}_{t-P} + \mathbf{e}_t
\label{eq:VARexpression}
\end{equation}
and $\mathbf{e}_t$ is uncorrelated process, here assumed to be Gaussian with unit covariance matrix. 
The positive definiteness of $R_Z$ in eq.(\ref{eq:AllCrossCorrM}) is important because it determines the stationarity of the VAR$_K(P)$ Gaussian process $\{Z_{1,t},\ldots,Z_{K,t}\}$. In terms of $A_1,\ldots,A_P$, VAR$_K(P)$ is stationary when the roots of the reverse characteristic polynomial $|I_K - A_1 \mathbf{z} - A_2 \mathbf{z}^2 - \cdots - A_P \mathbf{z}^P|=0$ lie outside the unit circle in the complex plane, or equivalently the eigenvalues of 
\[
\left[ \begin{array}{ccccc}
A_1 & A_2 & \cdots & A_{P-1} & A_P \\
I_K & 0 & \cdots & 0 & 0 \\
\vdots & \vdots & \vdots & \vdots & \vdots \\
0 & 0 & \cdots & I_K & 0 
\end{array} \right] 
\]
are all smaller than one in modulus, where $I_K$ is the unit matrix. This condition is fulfilled by the condition of positive semidefiniteness of the full correlation matrix of $R_Z$ in eq.(\ref{eq:AllCrossCorrM}).

Having a stationary VAR$_K(P)$ Gaussian process, we generate a stationary Gaussian multivariate time series $\{z^*_{i,t}\}_{t=1}^{N}$, $i=1,\ldots,K$ from eq.(\ref{eq:VARexpression}). To transform the Gaussian $\{z^*_{i,t}\}_{t=1}^{N}$ to possess the given marginals in step~\ref{step:VSTAP4} of the typical realization approach in Sec.~\ref{subsec:VSTAPsolution}, we can use either the marginal transform in eq.(\ref{eq:ztox}) or the linear piece-wise approximation. The former gives $\{x^*_{i,t}\}_{t=1}^{N}$ with exactly the same marginals and possibly some inaccuracy in the lagged correlations inherited by the inaccuracy of the linear piece-wise fit, while the latter loses some accuracy in matching the marginals but gains more accuracy in matching the lagged cross-correlations. The choice depends on the application. For example, for the randomization test for nonlinearity (where also we have $N=n$), we would choose the first approach to assure that the randomized time series contain exactly the same values as the original time series. We apply the latter approach in our simulations. 

%%%%%%%%%%%%%%%%%%%%%%%%%%%%%%
\section{Simulations and Results}
\label{sec:Simulations}
%%%%%%%%%%%%%%%%%%%%%%%%%%%%%%

We show the efficiency of VSTAP in generating multivariate time series that match given non-Gaussian marginals and various correlation structures. In all simulations we set the number of breakpoints in the piecewise approximation to 20. A pilot study on smaller number of breakpoints showed that the decreased accuracy of the piece-wise linear fit does not affect much the solution for $r_{Z_{i},Z_{j}}(\tau)$. So, for very small time series, a smaller number of breakpoints can also be used. On the other hand, our simulations on larger numbers of breakpoints showed insignificant improvement in the accuracy of the piece-wise linear fit and the estimated $r_{Z_{i},Z_{j}}(\tau)$. For the accuracy tolerance, we use the absolute error rather than a relative error and set $\epsilon = 0.00001$. For a generated realization of a process, VSTAP runs for the sample marginal distributions $F_{X_i}$ and the sample lagged correlations $r_{X_i,X_j}(\tau)$. Using the approach that matches exactly the marginals in step~\ref{step:VSTAP4} of Sec.~\ref{subsec:VSTAPsolution}, we establish that the marginals in $\{x^*_{i,t}\}_{t=1}^{n}$ coincide with those of the given time series $\{x_{i,t}\}_{t=1}^{n}$ (we use $n=N$ in the simulations). Thus only results for $r_{X_i,X_j}(\tau)$ are shown.  

\begin{figure}[h!]
\centering
\centerline{\hbox{\includegraphics[width=60mm]{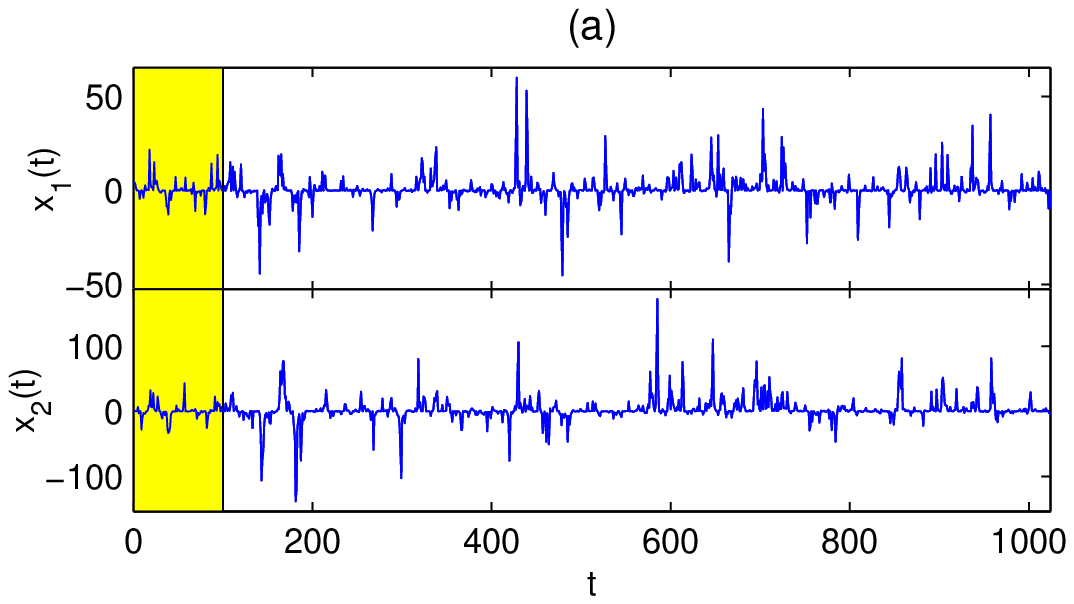}
\includegraphics[width=40mm]{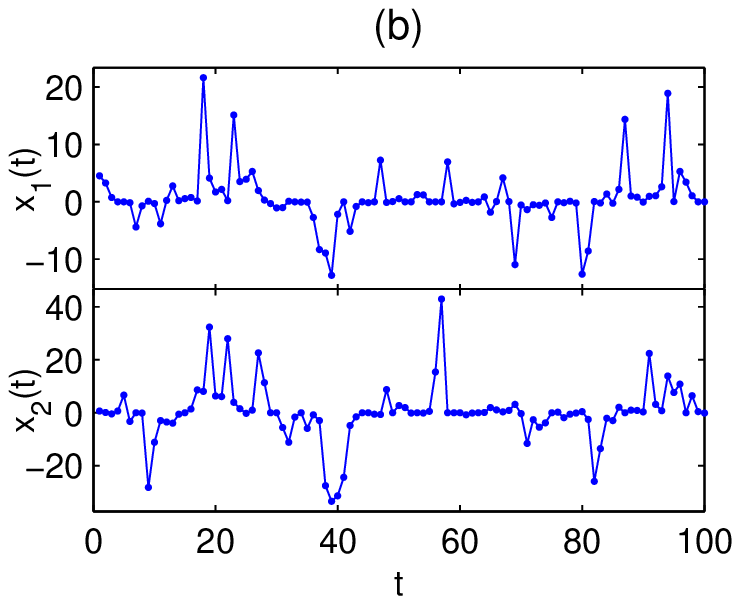}
\includegraphics[width=40mm]{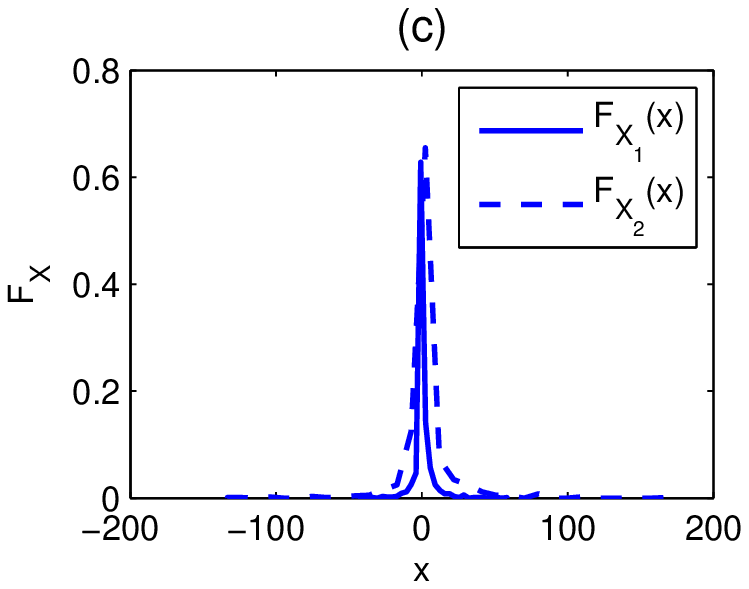}}}
\centerline{\hbox{\includegraphics[width=60mm]{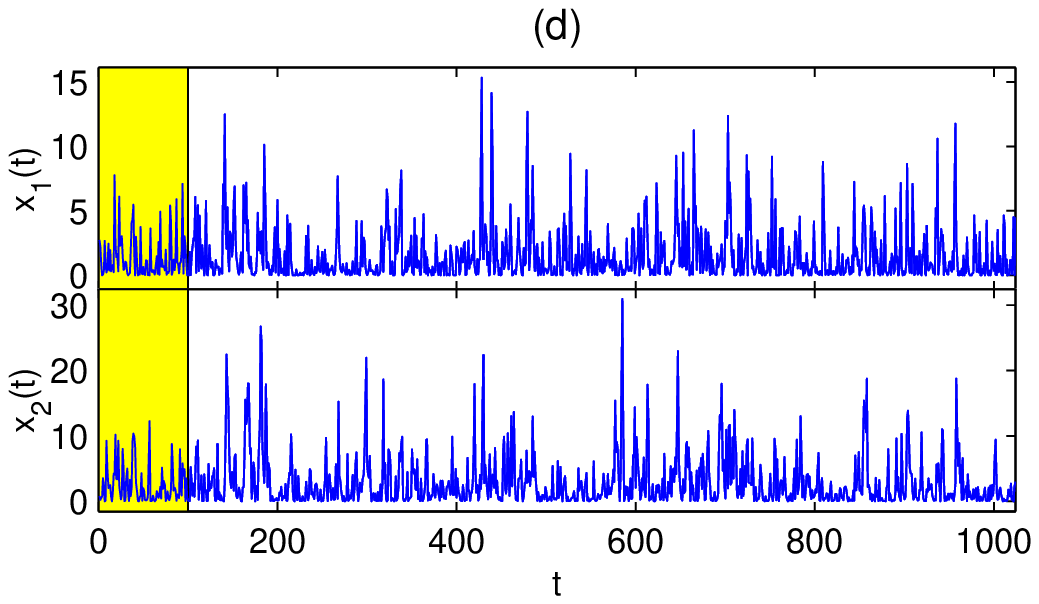}
\includegraphics[width=40mm]{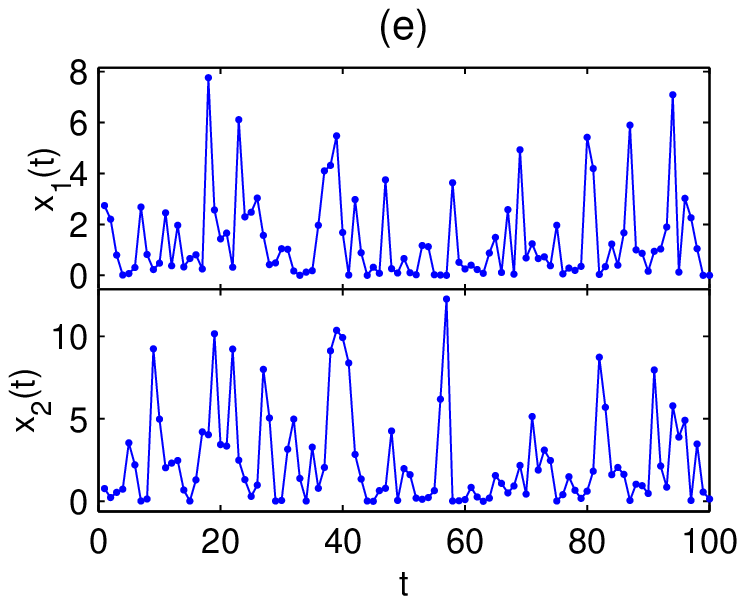}
\includegraphics[width=40mm]{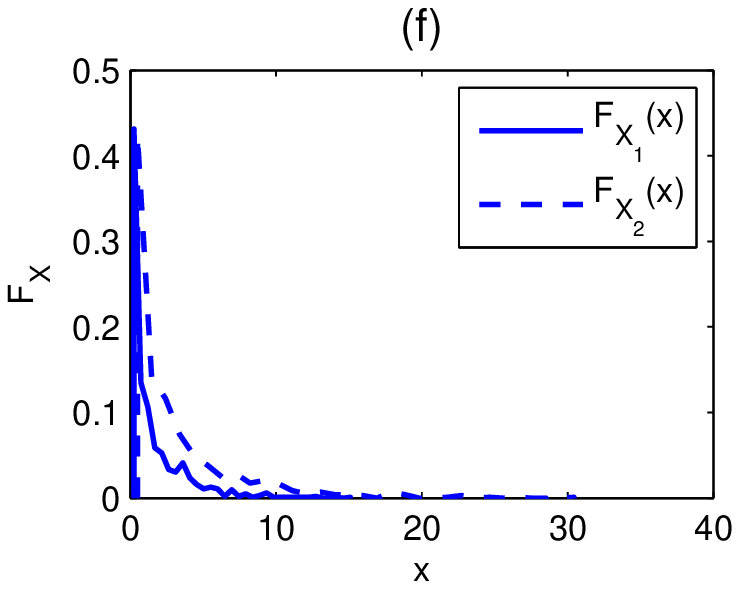}}}
\caption{The time series for monotonic ($a=3$, first row) and non-monotonic ($a=2$, second row) marginal transform of the Gaussian VAR$_2(2)$ generating process. (a) and (d) The time history plot for the two variables $X_1$ and $X_2$ for $a=3$ and $a=2$, respectively. (b) and (e) Blow up of the time windows as indicated in (a) and (d), respectively. (c) and (f) The histograms of the  time series for $X_1$ and $X_2$ in (a) and (d), respectively.}
 \label{fig:VAR22tshist}
\end{figure}

We consider different stationary VAR processes. We start with the Gaussian VAR$_2(2)$ generating process ($K=2$, $P=2$)  
\[
\mathbf{s}_t = \left[\begin{array}{c} s_{1,t} \\ s_{2,t} \end{array} \right] =
\left[\begin{array}{c} 0.02 \\ 0.03 \end{array} \right] + \left[\begin{array}{cc} 0.5 & 0.1 \\ 0.4 & 0.5 \end{array} \right] \mathbf{s}_{t-1} + \left[\begin{array}{cc} 0 & 0 \\ 0.25 & 0 \end{array} \right] \mathbf{s}_{t-2} + 
\mathbf{e}_t
\]
where the input white noise vector $\mathbf{e}_t=[e_{1,t}, e_{2,t}]^{\prime}$ is uncorrelated and has component variances $\sigma_{e_1}^2=0.09$ and $\sigma_{e_2}^2=0.04$ \citep[p.17]{Luetkepohl05}.
The observed time series $\{\mathbf{x}_t\}_{t=1}^n$, where $\mathbf{x}_t = [x_{1,t}, x_{2,t}]^{\prime}$, has altered (non-Gaussian) marginals, given as $x_{1,t}=s_{1,t}^a$, $x_{2,t}=s_{2,t}^a$, and we set $a=3$ and $a=2$ to have a monotonic and a non-monotonic marginal transform, respectively. An example of realizations of the processes for $a=3$ and $a=2$ are shown in Fig.~\ref{fig:VAR22tshist}. For $a=3$, the marginal distributions for both $X_1$ and $X_2$ have large kurtosis, while for $a=2$ both  marginal distributions are strongly right skewed.

\begin{figure}[h!]
\centering
\centerline{\hbox{\includegraphics[width=60mm]{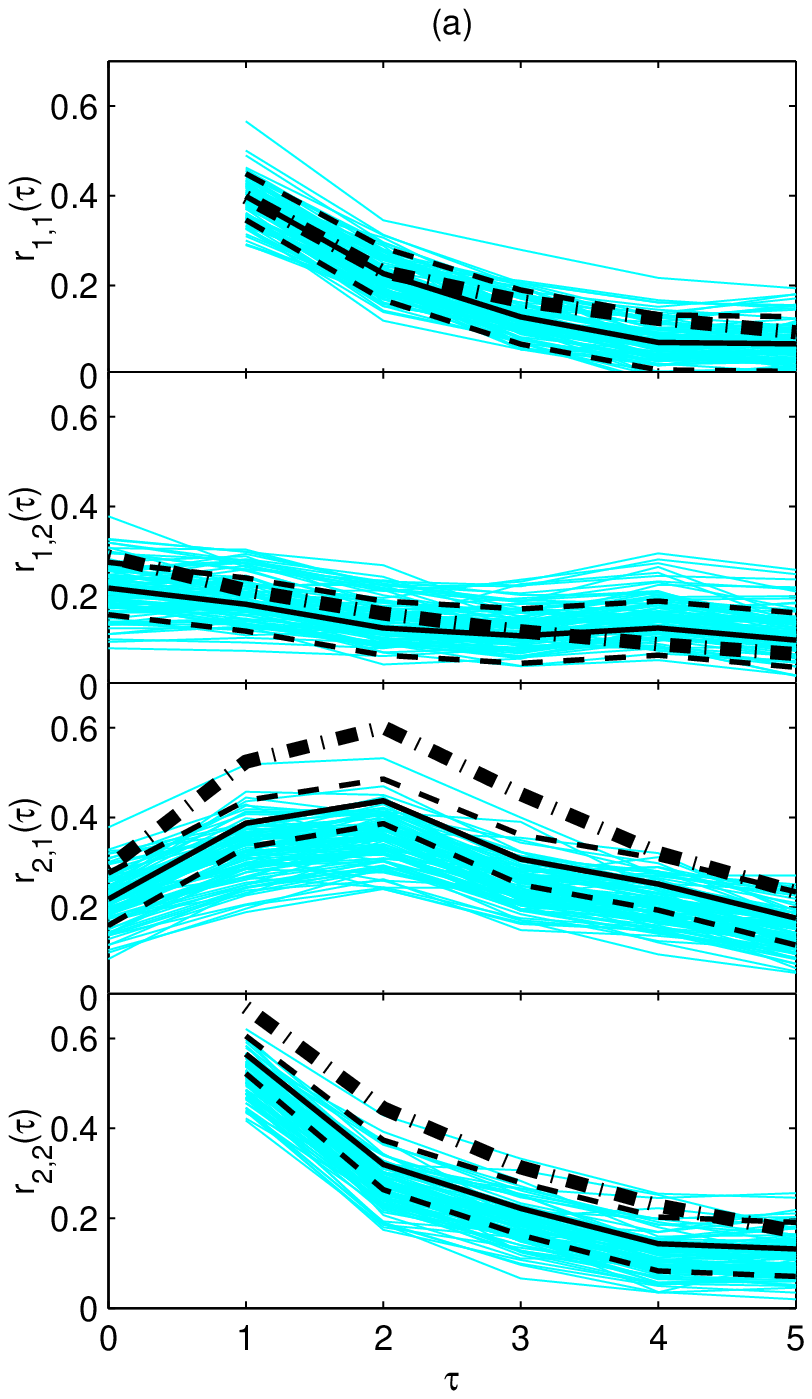}
\includegraphics[width=60mm]{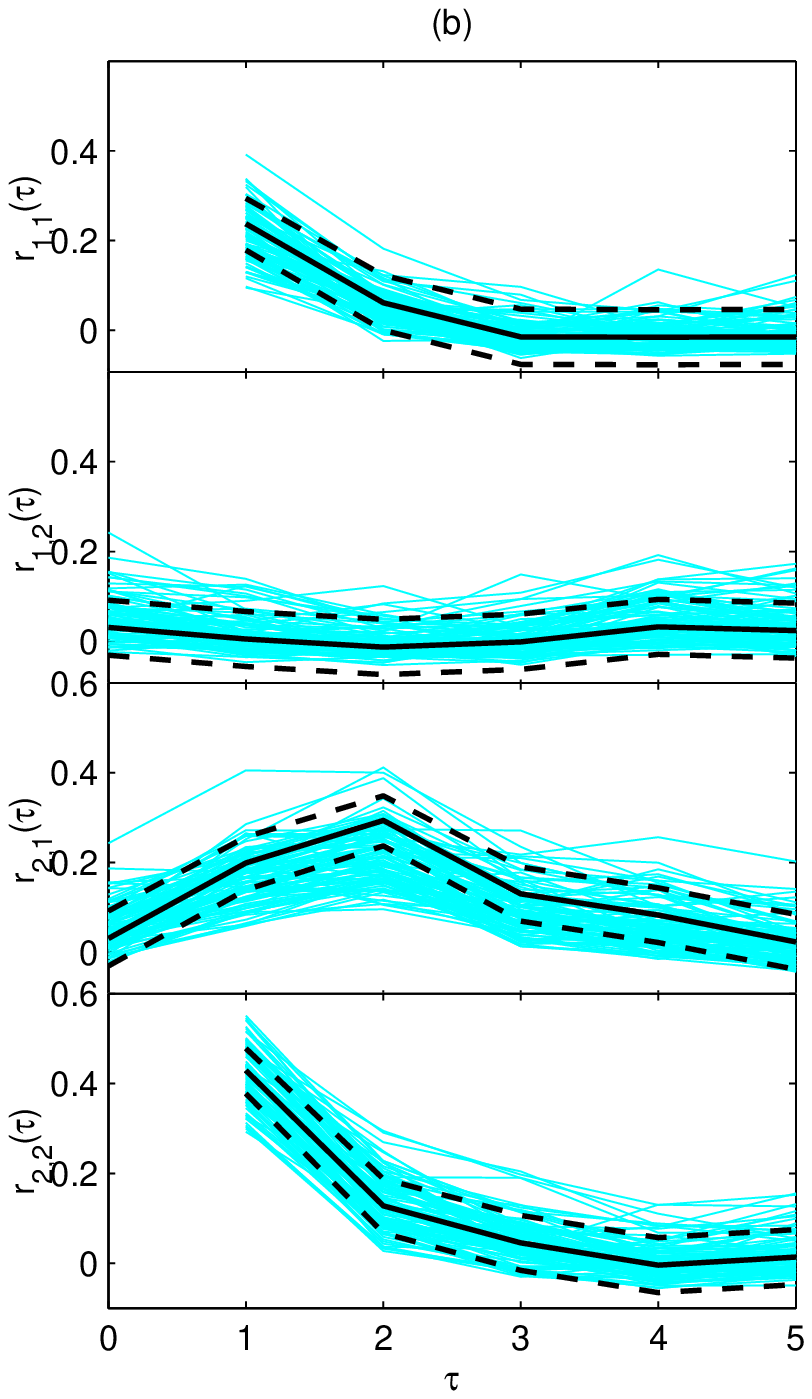}}}
\caption{The match of $r_{X_i,X_j}(\tau)$, $i,j=1,2$, using VSTAP for monotonic ($a=3$, in (a)) and non-monotonic ($a=2$, in (b)) marginal transform of the Gaussian VAR$_2(2)$ generating process. The black line is for the given $r_{X_i,X_j}(\tau)$ and the grey (cyan online) lines are for $r_{X_i^*,X_j^*}(\tau)$ from 100 realizations of VSTAP ($N=1024$). The two dashed black lines denote the 95\% Fisher confidence intervals of $r_{X_i,X_j}(\tau)$. In (a), the expected correlation from the cubic marginal transform is displayed by thick stippled line.}
 \label{fig:VAR2_2_P}
\end{figure}

The match of $r_{X_i,X_j}(\tau)$ with VSTAP for the monotonic and non-monotonic marginal transform can be seen in Fig.~\ref{fig:VAR2_2_P}. 
The 100 generated multivariate time series $\{x^*_{i,t}\}_{t=1}^{1024}$ have sample lagged correlations $r_{X_i^*,X_j^*}(\tau)$ that spread around the given $r_{X_i,X_j}(\tau)$, i.e. $r_{X_i,X_j}(\tau)$ is within the distribution of $r_{X_i^*,X_j^*}(\tau)$. The spread of $r_{X_i^*,X_j^*}(\tau)$ is at the level of the spread expected for sample Gaussian correlation of the same $N$, as indicated in Fig.~\ref{fig:VAR2_2_P} by the dashed black lines denoting the 95\% Fisher confidence intervals of $r_{X_i,X_j}(\tau)$. 

For the monotonic transform, it is possible to compare the given sample correlation and the VSTAP correlation with the theoretic correlation for the monotonically transformed Gaussian VAR$_2(2)$ process. For each Gaussian lagged correlation $\rho_{S_i,S_j}(\tau)$ of the original Gaussian VAR$_2(2)$ obtained from the coefficients of VAR$_2(2)$ through the Yule-Walker equations, the transform $x_{i,t}=s_{i,t}^3$, $i=1,2$, determines the correlation transform $\rho_{X_i,X_j}(\tau) = 2 \rho_{S_i,S_j}(\tau)^3/5 + 3\rho_{S_i,S_j}(\tau)/5$ \citep{Kugiumtzis10}. It turns out that both the sample and VSTAP lagged correlations are close to the theoretic lagged correlations for $(X_1,X_1)$ and $(X_1,X_2)$ but differ for $(X_2,X_1)$ and $(X_2,X_2)$ (see Fig.~\ref{fig:VAR2_2_P}a). 

\begin{figure}[h!]
\centering
\centerline{\hbox{\includegraphics[width=60mm]{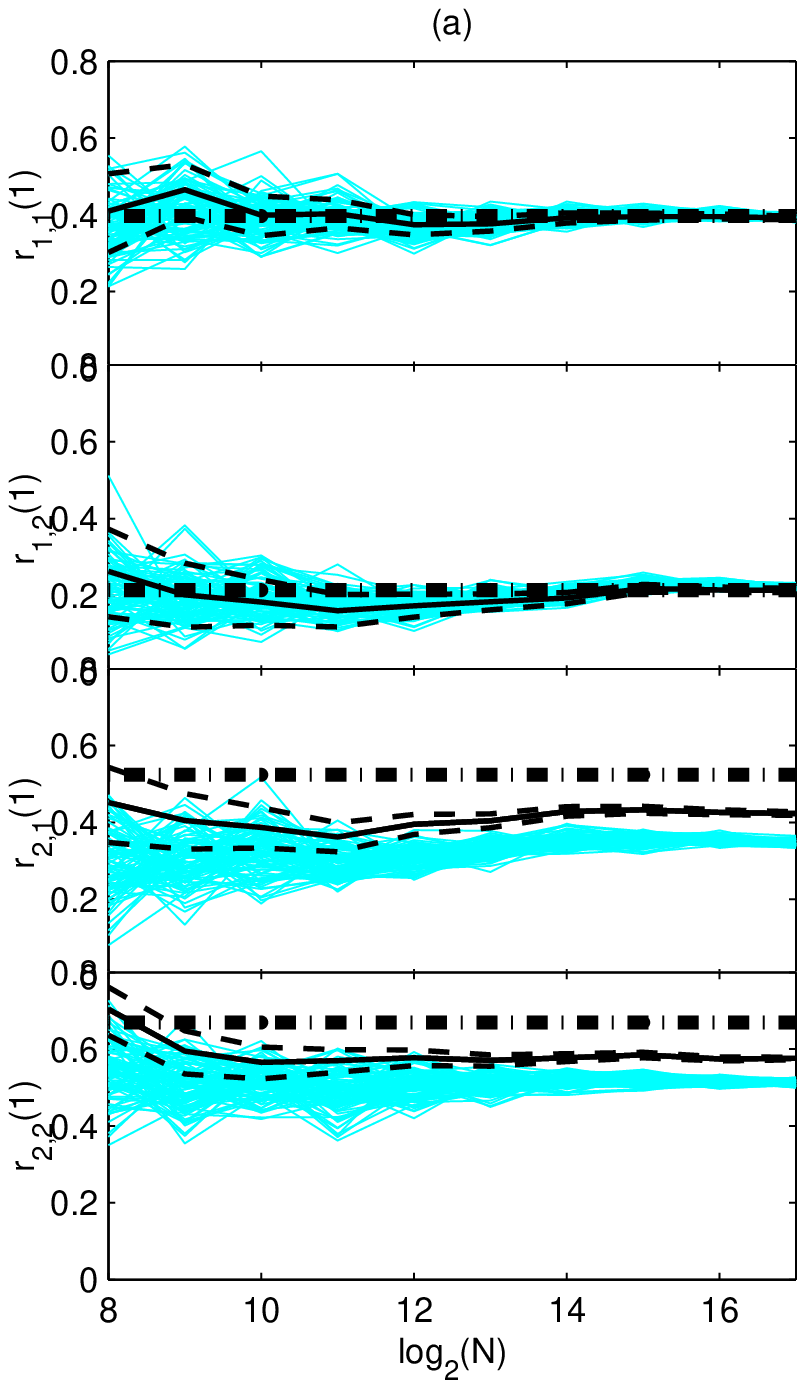}
\includegraphics[width=60mm]{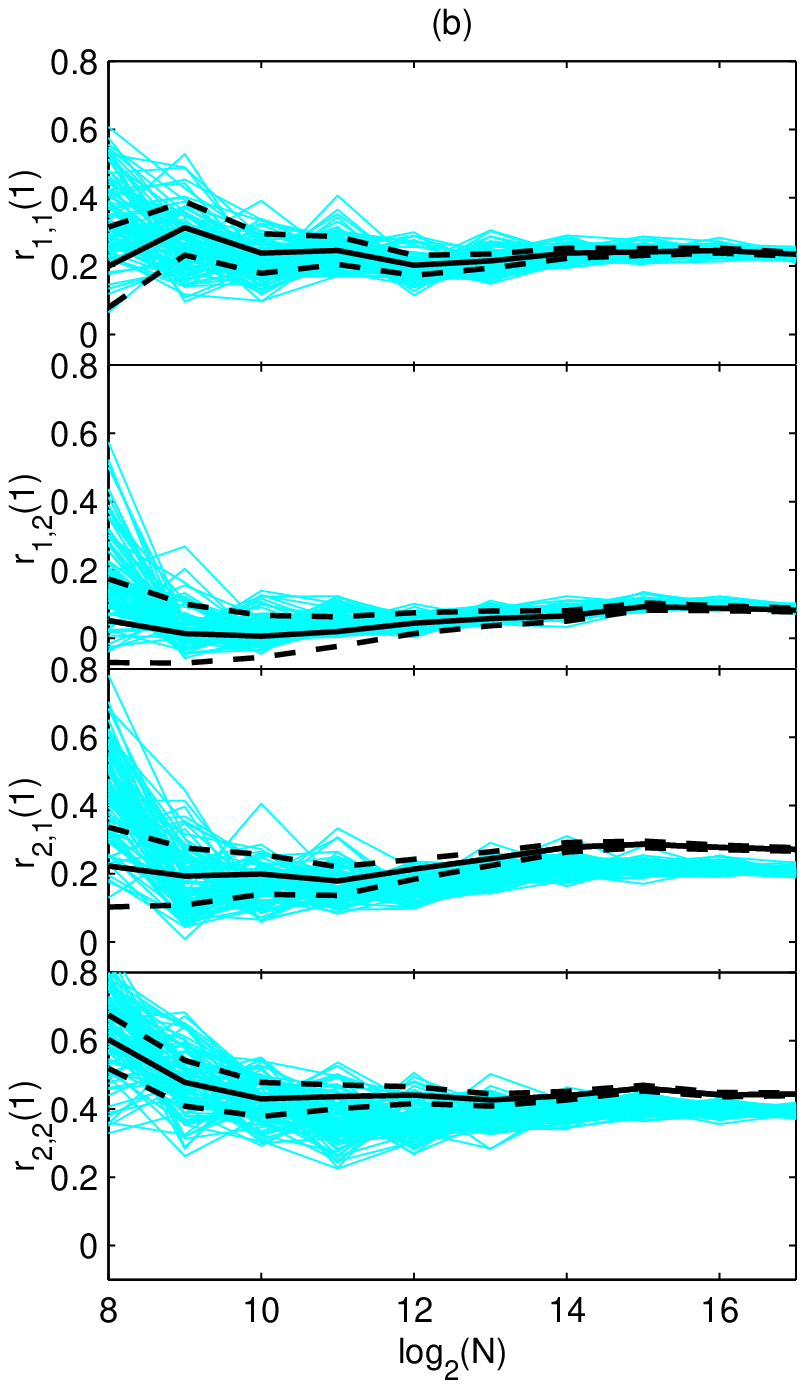}}}
\caption{The match of $r_{X_i,X_j}(1)$, $i,j=1,2$, against the time series length $N$ (logarithmic scale) using VSTAP for monotonic ($a=3$, in (a)) and non-monotonic ($a=2$, in (b)) marginal transform of the Gaussian VAR$_2(2)$ generating process. The black line is for $r_{X_i,X_j}(1)$ and the grey (cyan online) lines are for $r_{X_i^*,X_j^*}(1)$ from 100 realizations of VSTAP. The dashed black lines denote the 95\% Fisher confidence intervals of $\rho_{X_i,X_j}(1)$. In (a), the expected correlation from the cubic marginal transform is displayed by thick stippled line.}
 \label{fig:VAR2_2_n}
\end{figure}

The mismatch of theoretical and sample correlation is more visible when the sample size $N$ increases and the spread decreases, as shown in Fig.~\ref{fig:VAR2_2_n} for $\tau=1$. 
It is clearly shown in Fig.~\ref{fig:VAR2_2_n}a that the mismatch when $a=3$ occurs for $\rho_{X_2,X_1}(1)$ and $\rho_{X_2,X_2}(1)$. For these cases, one can observe also difference in the sample correlations $r_{X_2,X_1}(1)$ and $r_{X_2,X_2}(1)$ and the VSTAP correlations $r_{X_2^*,X_1^*}(1)$ and $r_{X_2^*,X_2^*}(1)$, respectively, which is actually due to the bias in the estimation of the correlation matrix of the VAR process. 
Monte Carlo simulations on realizations of the Gaussian VAR$_2(2)$ process, without applying marginal transform and VSTAP, showed that the bias occurs even when the process is Gaussian. This bias is thus passed also to the VSTAP estimation of the given sample lagged correlations. For the monotonic marginal transform, we note that the mismatch of VSTAP for large $N$ occurs only when there is bias, i.e. the sample correlation $r_{X_i,X_j}(1)$ differs from the theoretic correlation $\rho_{X_i,X_j}(1)$ (see Fig.~\ref{fig:VAR2_2_n}a).  

\begin{figure}[h!]
\centering
\centerline{\hbox{\includegraphics[width=60mm]{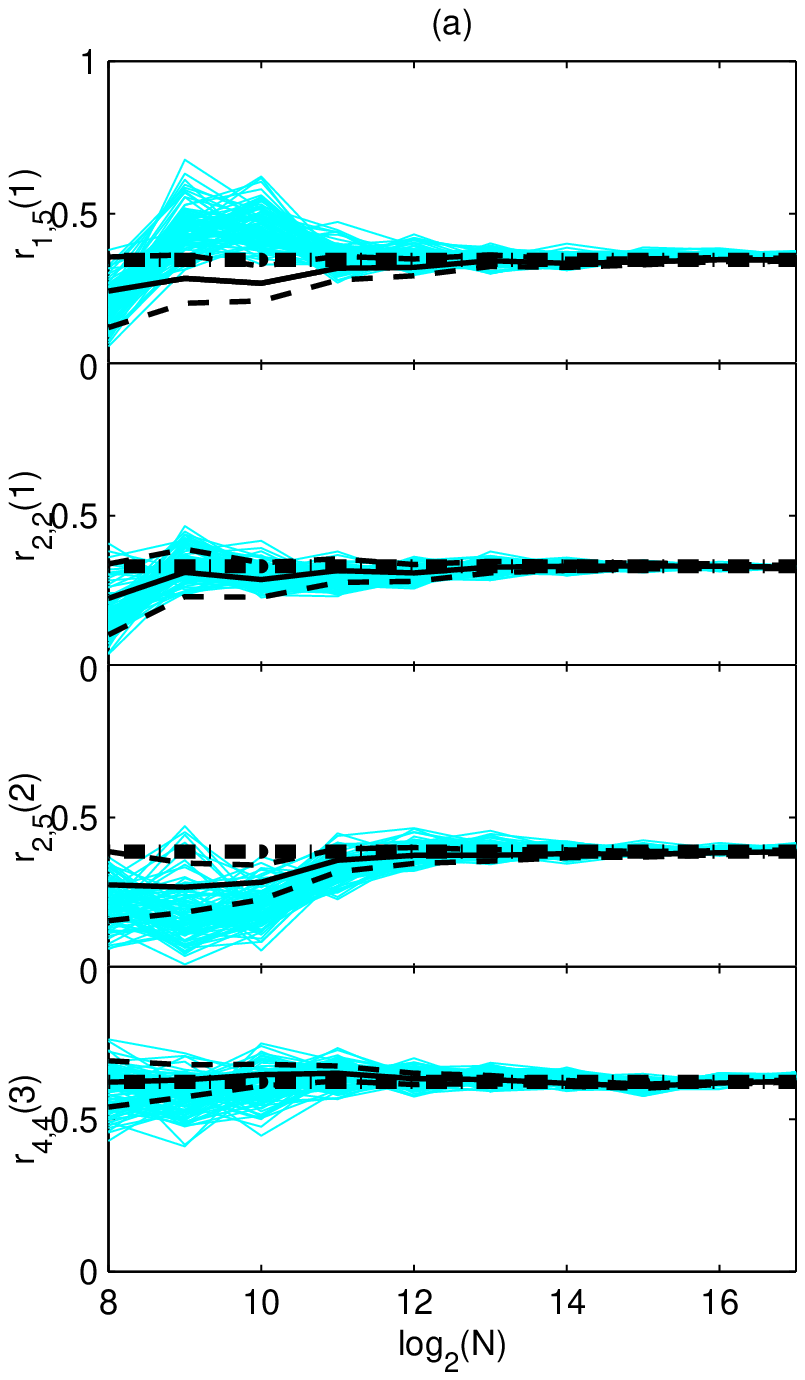}
\includegraphics[width=60mm]{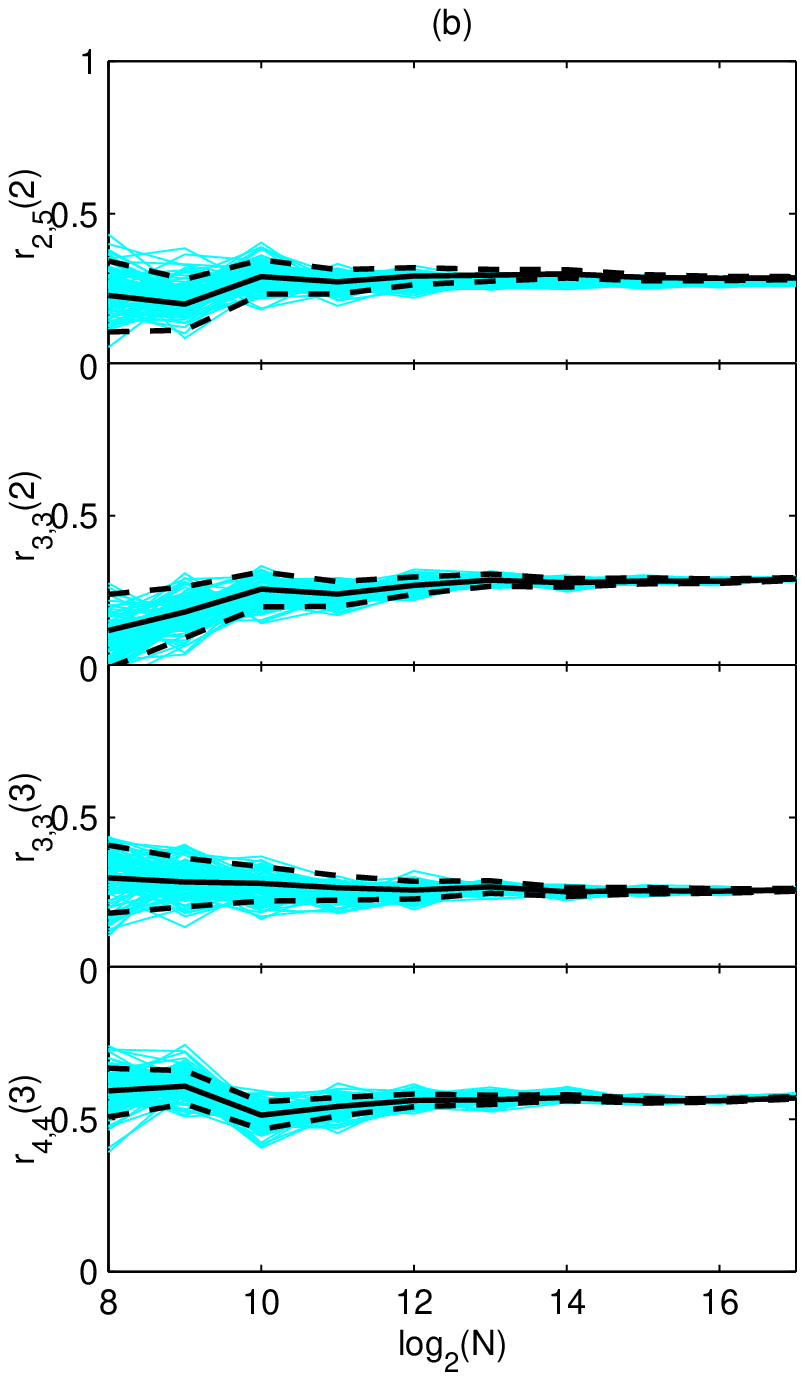}}}
\caption{The match of $r_{X_i,X_j}(\tau)$, $i,j=1,2$, against the time series length $N$ (logarithmic scale) using VSTAP for monotonic ($a=3$, first row) and non-monotonic ($a=2$, second row) marginal transform of the Gaussian VAR$_5(4)$ generating process. The black line is for $r_{X_i,X_j}(\tau)$ and the grey (cyan online) lines are for $r_{X_i^*,X_j^*}(\tau)$ from 100 realizations of VSTAP, for the four largest $r_{X_i,X_j}(\tau)$ found when $N=2^{17}$. The dashed black lines denote the 95\% Fisher confidence intervals of $r_{X_i,X_j}(\tau)$. In the first row the expected correlation from the cubic marginal transform is displayed by thick stippled line.}
 \label{fig:VAR5_4_n}
\end{figure}

The results on other systems showed a better estimation of the theoretical lagged correlations (for monotonic marginal transforms), and therefore also VSTAP matched better the given sample lagged correlations. For example, we made the same computations for a Gaussian VAR$_5(4)$ process undergoing the same monotonic and non-monotonic marginal transforms
\[
\begin{array}{ccc}
s_{1,t} & = & 0.4 s_{1,t-1}-0.5 s_{1,t-2}+0.4 s_{5,t-1}+e_{1,t} \\
s_{2,t} & = & 0.4 s_{2,t-1}-0.3 s_{1,t-4}+0.4 s_{5,t-2}+e_{2,t} \\
s_{3,t} & = & 0.5 s_{3,t-1}-0.7 s_{3,t-2}-0.3 s_{5,t-3}+e_{3,t} \\
s_{4,t} & = & 0.8 s_{4,t-3}+0.4 s_{1,t-2}+0.3 s_{2,t-2}+e_{4,t} \\
s_{5,t} & = & 0.7 s_{5,t-1}-0.5 s_{5,t-2}-0.4 s_{4,t-1}+e_{5,t} 
\end{array}
\]
where the input white noise vector has unit covariance matrix 
(the system was first introduced in \citep{Schelter06e}). 
VSTAP matched well the given sample lagged correlations, and even for large $N$ the ensemble of $r_{X_i^*,X_j^*}(\tau)$ was spread around $r_{X_i,X_j}(\tau)$ for almost all pairs $(X_i,X_j)$ and $\tau=0,\ldots,5$, as shown in Fig.~\ref{fig:VAR5_4_n} for the 4 largest $r_{X_i,X_j}(\tau)$ and the monotonic and non-monotonic transforms. 
The cases of $r_{X_i,X_j}(\tau)$ not falling in the distribution of $r_{X_i^*,X_j^*}(\tau)$ were all for very small $r_{X_i,X_j}(\tau)$ and the deviation was also small and could be observed only for very large $N$.

\begin{figure}[h!]
\centering
\centerline{\hbox{\includegraphics[width=60mm]{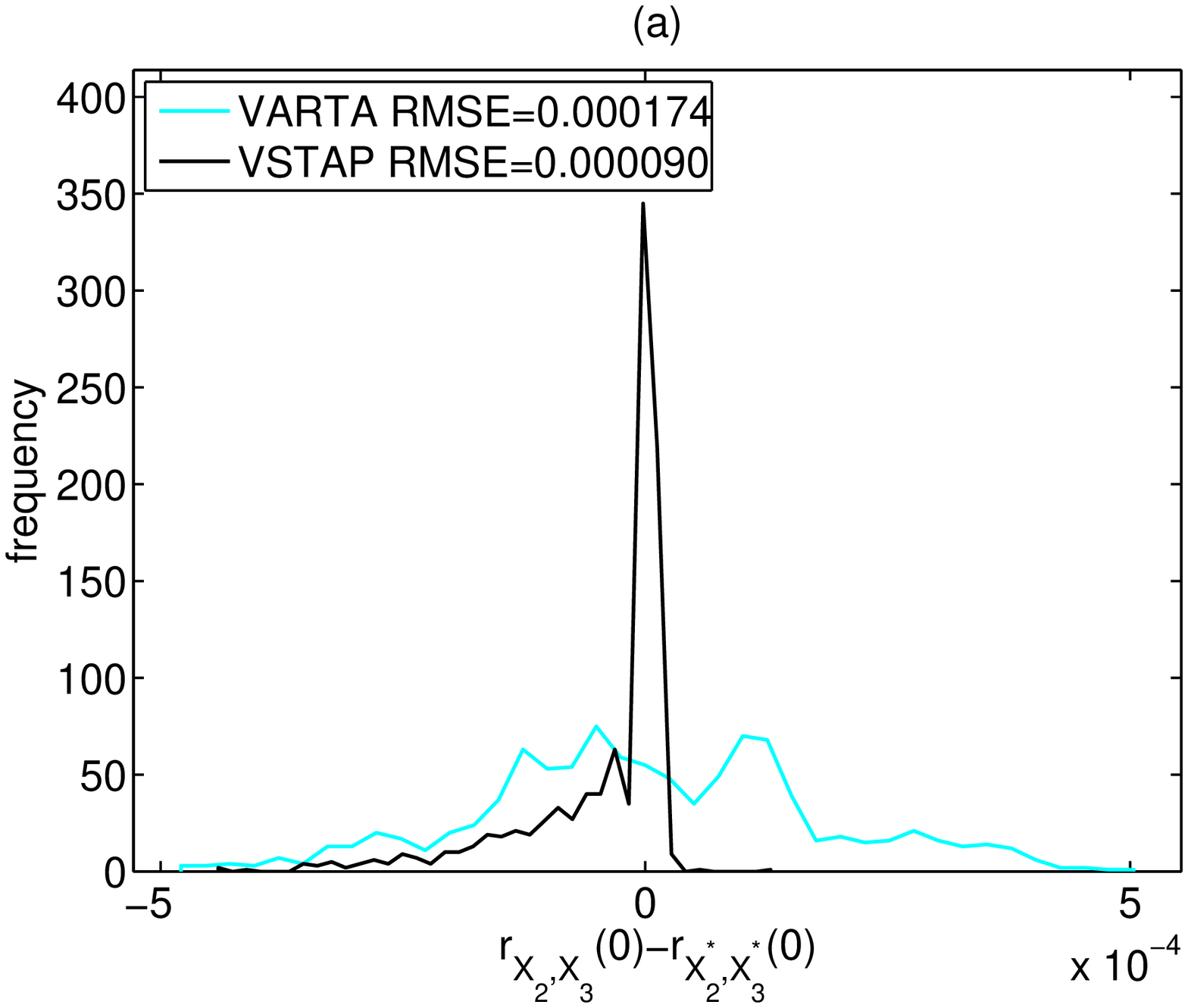}
\includegraphics[width=60mm]{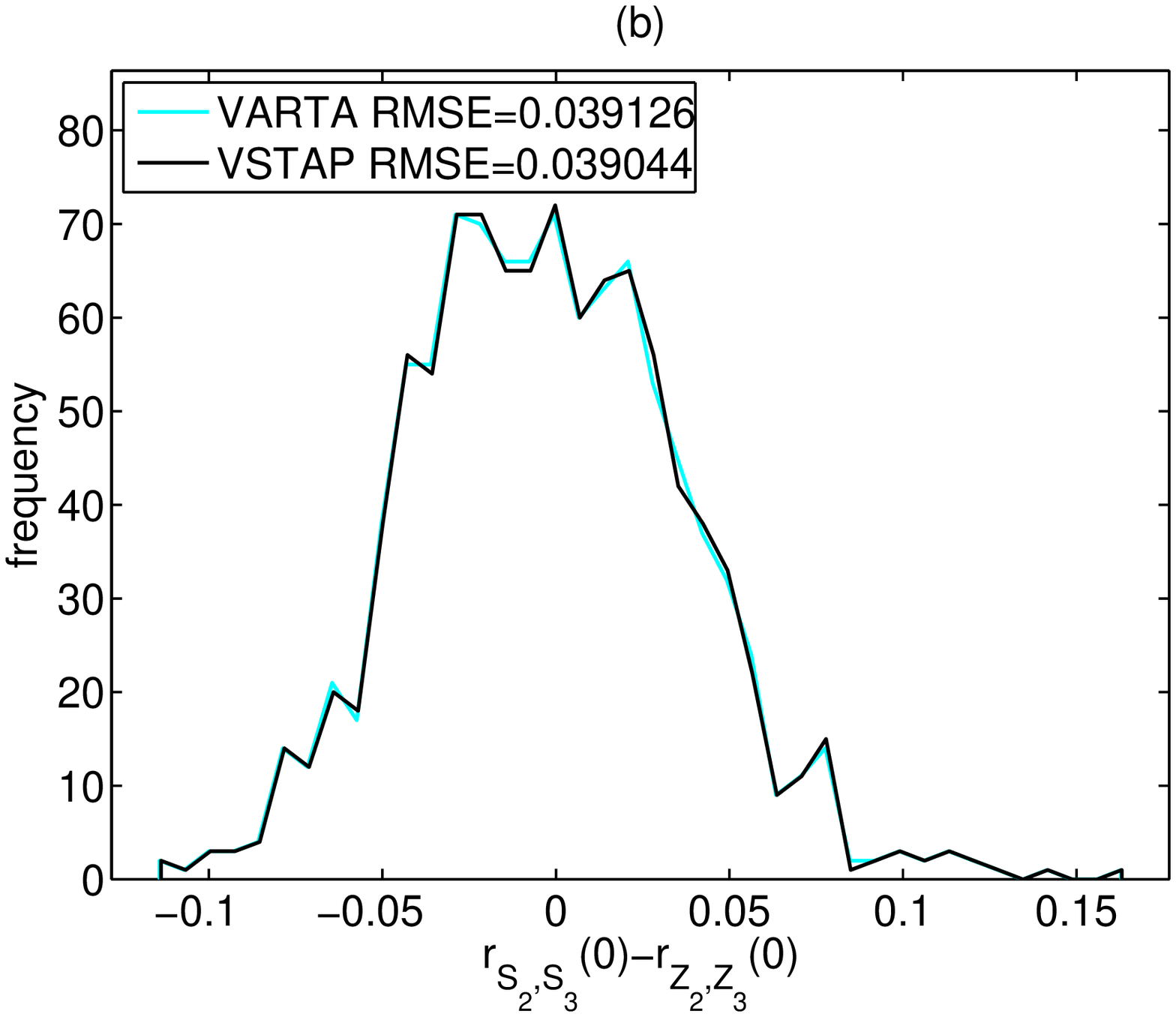}}}
\caption{(a) The histogram of the difference $r_{X_2,X_3}(0)-r_{X_2^*,X_3^*}(0)$ in 1000 realizations of $n=128$ for the example of three uniform variables obtained by VARTA (or NORTA) and VSTAP, as shown in the legend, where also the corresponding root mean square error is indicated. (b) The same as in (a) but for the Gaussian correlation coefficient $r_{S_2,S_3}(0)$.}
 \label{fig:NORTA1}
\end{figure}

Finally, we report some results on the computational efficiency of VSTAP and compare it to VARTA. We focus on the computation of single components of the correlation matrix and consider a simplified setting of no auto-correlation, which is equivalent to having samples of correlated variables. In particular, we consider the example of a non-feasible Gaussian correlation matrix for three uniform variables with $r_{X_1,X_2}(0)\!\!=\!\!-0.4$, $r_{X_1,X_3}(0)\!\!=\!\!0.2$ and $r_{X_2,X_3}(0)\!\!=\!\!0.8$, first reported in \cite{Li75}. The Gaussian correlation coefficients are given analytically from the expression $r_{S_1,S_2}=2\sin(\pi r_{X_1,X_2} / 6)$ as $r_{S_1,S_2}(0)\!\!=\!\!-0.4158$, $r_{S_1,S_3}(0)\!\!=\!\!0.2091$ and $r_{S_2,S_3}(0)\!\!=\!\!0.8135$. The corresponding correlation matrix is not positive semidefinite and applying eigenvalue correction we derive the closest positive semidefinite correlation matrix with components $r_{S_1,S_2}(0)\!\!=\!\!-0.4122$, $r_{S_1,S_3}(0)\!\!=\!\!0.2062$ and $r_{S_2,S_3}(0)\!\!=\!\!0.8065$. Using these correlation coefficients, 1000 multivariate Gaussian samples are generated and transformed to uniform marginals applying the Gaussian cumulative density function. Since the marginal transform is monotonic we can evaluate VSTAP and VARTA (for this setting this is actually equivalent to the algorithm of normal to anything (NORTA) \cite{Chen01}) in matching $r_{X_i,X_j}(0)$ and $r_{S_i,S_j}(0)$. The results for one of the three variable pairs is given in Figure~\ref{fig:NORTA1}, and similar are the results for the other two pairs.
Both methods ran for the same accuracy level, given by a relative error of 0.001. Thus both methods match well $r_{X_i,X_j}(0)$, with $r_{X_i^*,X_j^*}(0)$ of VARTA spreading evenly around $r_{X_i,X_j}(0)$, and $r_{X_i^*,X_j^*}(0)$ of VSTAP being mostly concentrated at $r_{X_i,X_j}(0)$ and spread over at larger values, giving somewhat smaller root mean square error (RMSE). VSTAP gives also smaller RMSE in matching $r_{S_i,S_j}(0)$, where both methods have much smaller accuracy in approximating $r_{S_i,S_j}(0)$.

\begin{figure}[h!]
\centering
\centerline{\hbox{\includegraphics[width=60mm]{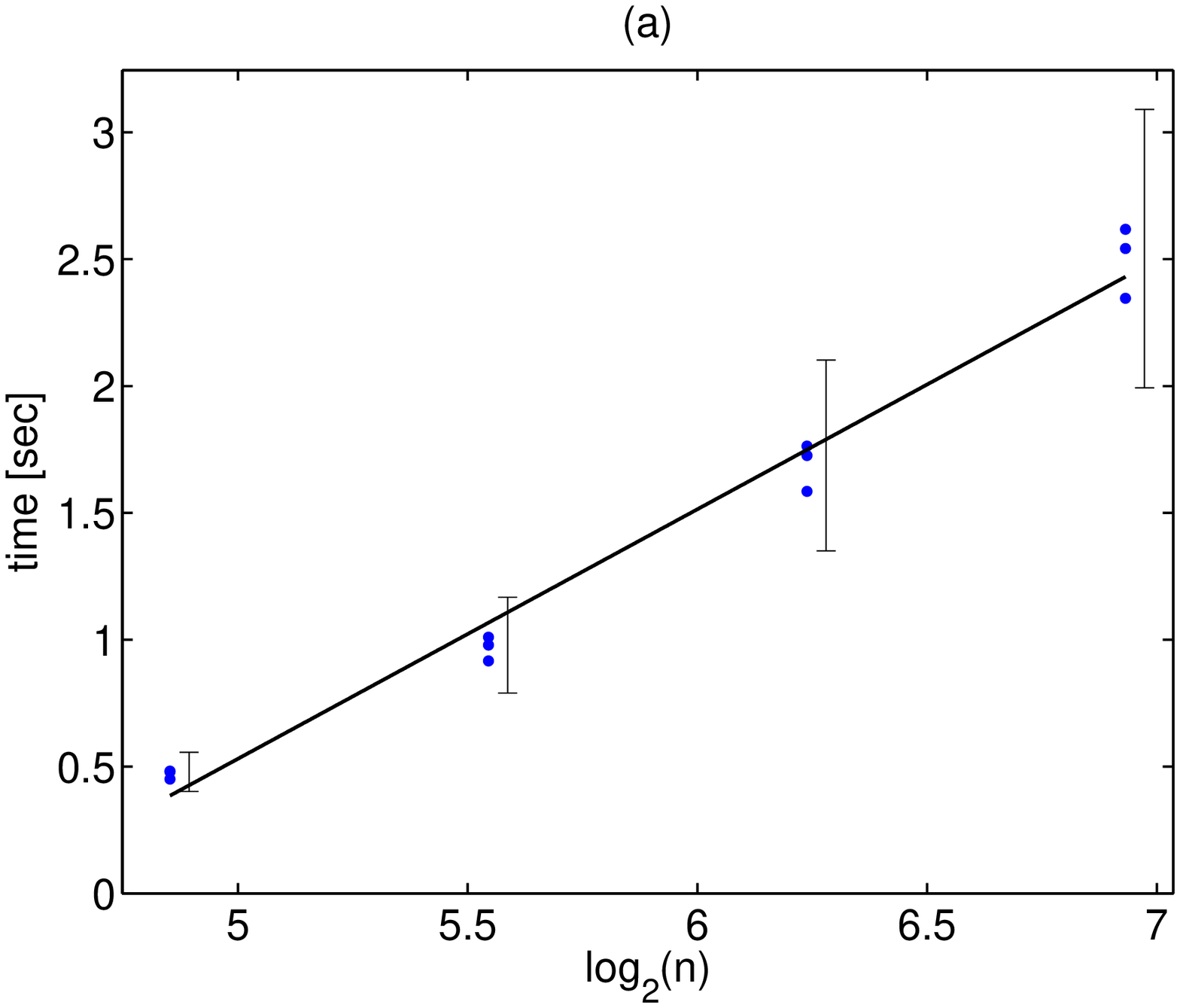}
\includegraphics[width=60mm]{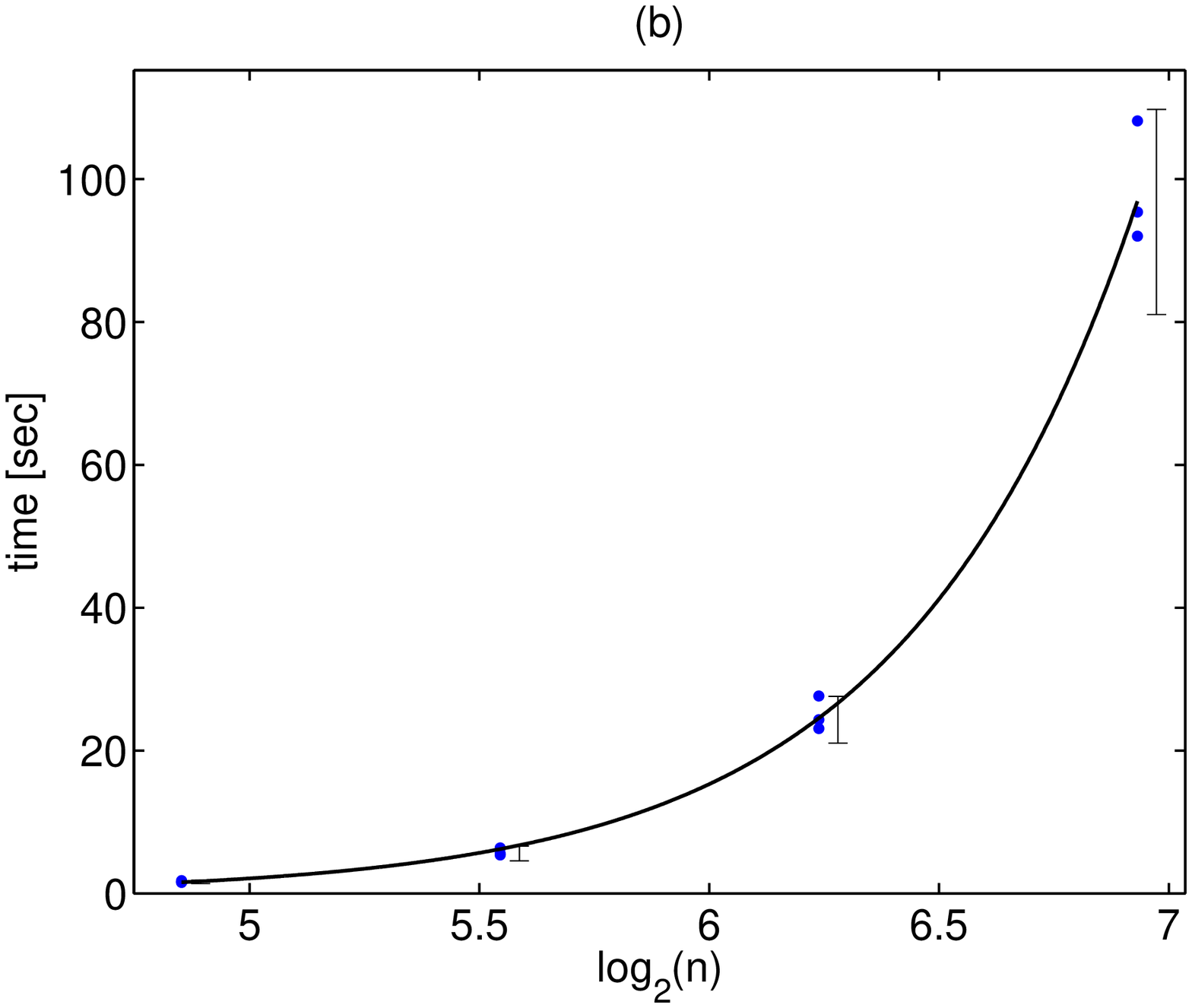}}}
\caption{(a) The time for the computation of $r_{X_i^*,X_j^*}(0)$ as a function of $n$, for the three pairs of the three uniform variables using VSTAP. For each $n$ the three drawn points correspond to the mean computation time for each of the three variable pairs from 1000 realizations. The error bar drawn for each $n$ regards the standard deviation of the computation time. The solid line is the best logarithmic fit. (b) The same as in (a) but for VARTA and the solid line is the best fit of a power of $n$.}
 \label{fig:NORTA2}
\end{figure}

While both VSTAP and VARTA attain the same accuracy level in approximating $r_{X_i,X_j}(0)$, VSTAP succeeds this much faster\footnote{The calculations were done on a PC with Intel Core i7 CPU 3.07GHz and 12GB RAM and for VSTAP the code was developed in Matlab while for VARTA the Fortran code in \url{http://users.iems.northwestern.edu/~nelsonb/ARTA} was used after slight modification of input/output.}. As shown in Figure~\ref{fig:NORTA2} for samples sizes $n=128,256,512,1024$, the computation time increases slowly with $n$ for VSTAP and fast with VARTA. 
In particular, for VSTAP the scaling is logarithmic, and the fit for the computation time $t$ in sec is $t=-4.39+0.98\ln(n)$, while for VARTA $t$ scales as a square of $n$, and the fit is $t=-9.16n^{1.98}$. The variance about the mean values in the 1000 realizations was relatively small, as shown by the error bars in Figure~\ref{fig:NORTA2} denoting the standard deviation. For the setting shown in Figure~\ref{fig:NORTA1} the mean computation time for VSTAP is 0.45 sec and for VARTA 1.55 sec with standard deviation 0.07 sec and 0.26 sec, respectively, indicating a significant difference even for small sample sizes. 

%%%%%%%%%%%%%%%%%%%%%%%%%%%%%%
\section{Conclusion}
\label{sec:Conclusion}
%%%%%%%%%%%%%%%%%%%%%%%%%%%%%%

The proposed method VSTAP can generate multivariate time series of arbitrary length with any given marginals and correlation structure, provided that the marginal distributions are continuous and the lagged correlation matrix is feasible. The general use of VSTAP lies in the linear piecewise approximation of the marginal transform from Gaussian to the given marginal, which allows for a closed form solution for the correlation transform from each component of the Gaussian lagged correlation matrix to the respective component of the given lagged correlation matrix. Thus any continuous marginal distribution, e.g. multi-modal or strongly skewed, can be sufficiently approximated at an accuracy depending on the number of breakpoints for the piecewise function. For all practical purposes, the accuracy converges with the number of breakpoints reaching the level of about 20, so that sufficient approximation of the sample marginal distribution is always obtained unless there are very few observed points, i.e. the time series is very short not allowing for the use of a sufficient number of breakpoints. Moreover, making use of the statistics of the joint doubly truncated Gaussian distribution, we could reach an analytic expression for the correlation transform. This allows for a straightforward and stable solution.

The VSTAP algorithm is also time effective, as the iterative scheme makes computation of a closed form expression for the correlation transform. This is to be compared to the numerical solution of the double integral form of the correlation transform used in the VARTA approach. We demonstrated with a simple example that piece-wise approximation results in much faster computations of the solution than numerical integration without any loss in accuracy. Still this may depend on the numerical integration scheme, which we did not investigate. So, besides the insight onto the correlation transform from Gaussian to target correlation provided by the closed-form approximation of the transform, a practical advantage of VSTAP is the derivation of the solution for the Gaussian correlation without the need of a time consuming two-dimensional numerical integration. 

In VSTAP, we treated the problem of obtaining proper correlation matrices. Simple eigenvalue correction to make the correlation matrix positive semi-definite is not directly applicable to the lagged correlation matrix as it contains repeated entries. We introduced an iterative two-stage procedure that turned out to render positive definiteness in just few steps. 

By construction VSTAP matches exactly the given marginals. In all simulations with different VAR processes, VSTAP could also match well the auto- and cross-correlations for a sufficiently large number of lags when the length of the generated multivariate time series was up to moderately large, say up to about 4000. For larger lengths, some deviation could be observed for some specific auto- and cross-correlations, which however were rare and only in some of the studied systems (besides the presented simulation results for two systems, a number of other VAR systems of varying structure were tested). Thus for most practical purposes VSTAP generates proper multivariate time series that can be used for randomization tests or in stochastic simulation. 

We have considered marginal transforms that deviate a lot from Gaussian and many different correlation structures. However, in our simulations, we have not encountered non-feasible correlation matrices as VSTAP would always provide sufficient solution, eventually after rendering positive definiteness of the lagged correlation matrix.  
It is therefore our intention to test VSTAP on special cases with more extreme marginals, e.g. positively skewed with a peak at zero, and stronger auto- and cross- correlations, e.g. typically expected from oscillating time series.  
The latter are often met in applications of the randomization test for nonlinearity, which is not discussed here but truly it has been the main motivation for this work. 
We leave this discussion and comparison to frequency-based methods, such as IAAFT for multivariate time series, including also time series from nonlinear dynamical systems, to future work.

%\bibliography{c:/MyFiles/Papers/LaTeX/StatsReferences}

\end{document}